\documentclass[conference]{IEEEtran}
\IEEEoverridecommandlockouts
\usepackage{cite}
\usepackage{amsmath,amssymb,amsfonts}
\usepackage{algorithmic}
\usepackage{graphicx}
\usepackage{textcomp}
\usepackage{xcolor}
\usepackage{float}
\usepackage{hyperref}
\usepackage{tikzit}
\usepackage[colorinlistoftodos]{todonotes}

\tikzstyle{blueRec}=[fill={rgb,255: red,160; green,203; blue,255}, draw=none, shape=rectangle, tikzit shape=rectangle, tikzit fill={rgb,255: red,160; green,203; blue,255}]
\tikzstyle{blueLargeRec}=[fill={rgb,255: red,188; green,226; blue,255}, draw=black, shape=rectangle, minimum width=84pt, minimum height=16pt, line width=1.pt]
\tikzstyle{greenLargeRec}=[fill={rgb,255: red,210; green,255; blue,199}, draw=black, shape=rectangle, minimum width=84pt, minimum height=16pt, line width=1.pt]
\tikzstyle{redLargeRec}=[fill={rgb,255: red,255; green,191; blue,191}, draw=black, shape=rectangle, minimum width=84pt, minimum height=16pt, line width=1.pt]
\tikzstyle{yellowLargRec}=[fill={rgb,255: red,255; green,250; blue,189}, draw=black, shape=rectangle, minimum width=84pt, minimum height=16pt, line width=1.pt]
\tikzstyle{etc}=[fill=black, draw=black, shape=circle, minimum width=2pt, minimum height=2pt]

\tikzstyle{fillBlue}=[-, fill={rgb,255: red,167; green,225; blue,255}]
\tikzstyle{fillGreen}=[-, fill={rgb,255: red,201; green,255; blue,189}]
\tikzstyle{fillYellow}=[-, fill={rgb,255: red,255; green,253; blue,183}]
\tikzstyle{myEdgeArrows}=[->, line width=1.pt]
\tikzstyle{myEdge}=[-, line width=1.pt]
\tikzstyle{new edge style 0}=[-, fill=white]
\tikzstyle{myEdge2}=[<-, line width=1.]
\tikzstyle{dataflow}=[draw={rgb,255: red,191; green,0; blue,64}, ->, line width=5pt]
\tikzstyle{dataflow2}=[draw={rgb,255: red,191; green,0; blue,64}, line width=5pt]
\tikzstyle{sync}=[-, draw=red, line width=2pt]

\def\BibTeX{{\rm B\kern-.05em{\sc i\kern-.025em b}\kern-.08em
    T\kern-.1667em\lower.7ex\hbox{E}\kern-.125emX}}
\begin{document}

\title{In-Situ Techniques on GPU-Accelerated Data-Intensive Applications}


\author{
\IEEEauthorblockN{Yi Ju \IEEEauthorrefmark{1}, Mingshuai Li \IEEEauthorrefmark{3}, Adalberto Perez \IEEEauthorrefmark{2}, Laura Bellentani \IEEEauthorrefmark{4}, Niclas Jansson \IEEEauthorrefmark{2}, \\ Stefano Markidis \IEEEauthorrefmark{2}, Philipp Schlatter \IEEEauthorrefmark{2}\IEEEauthorrefmark{5}, Erwin Laure \IEEEauthorrefmark{1}}
\IEEEauthorblockA{\IEEEauthorrefmark{1}\textit{Max Planck Computing and Data Facility}
 \{yi.ju, erwin.laure\}@mpcdf.mpg.de} 
\IEEEauthorblockA{\IEEEauthorrefmark{2}\textit{KTH Royal Institute of Technology}
 \{adperez, njansson, markidis, pschlatter\}@kth.se} 
\IEEEauthorblockA{\IEEEauthorrefmark{4}\textit{CINECA}
 l.bellentani@cineca.it \IEEEauthorrefmark{3}\textit{Technical University of Munich}
 mingshuai.li@tum.de}
\IEEEauthorblockA{\IEEEauthorrefmark{5}\textit{Friedrich-Alexander-Universität Erlangen-Nürnberg} philipp.schlatter@fau.de} 
}

\maketitle

\begin{abstract}

The computational power of High-Performance Computing (HPC) systems is constantly increasing, however, their input/output (IO) performance grows relatively slowly, and their storage capacity is also limited. This unbalance presents significant challenges for applications such as Molecular Dynamics (MD) and Computational Fluid Dynamics (CFD), which generate massive amounts of data for further visualization or analysis. At the same time, checkpointing is crucial for long runs on HPC clusters, due to limited walltimes and/or failures of system components, and typically requires the storage of large amount of data. 
Thus, restricted IO performance and storage capacity can lead to bottlenecks for the performance of full application workflows (as compared to computational kernels without IO). In-situ techniques, where data is further processed while still in memory rather to write it out over the I/O subsystem, can help to tackle these problems. In contrast to traditional post-processing methods, in-situ techniques can reduce or avoid the need to write or read data via the IO subsystem. They offer a promising approach for applications aiming to leverage the full power of large scale HPC systems. In-situ techniques can also be applied to hybrid computational nodes on HPC systems consisting of graphics processing units (GPUs) and central processing units (CPUs). On one node, the GPUs would have significant performance advantages over the CPUs. Therefore, current approaches for GPU-accelerated applications often focus on maximizing GPU usage, leaving CPUs underutilized. In-situ tasks using CPUs to perform data analysis or preprocess data concurrently to the running simulation, offer a possibility to improve this underutilization.

\end{abstract}


\section{Introduction}

\begin{figure*}[t]
\vspace{-2em}
	\centering
\resizebox{0.95\textwidth}{!}{%
	\begin{tikzpicture}
	\begin{pgfonlayer}{nodelayer}
		\node [style=blueLargeRec] (0) at (1.25, -4.75) {Application initialization};
		\node [style=blueLargeRec] (1) at (1.25, -7) {Application iteration};
		\node [style=blueLargeRec] (2) at (1.25, -9.75) {Application finalization};
		\node [style=greenLargeRec] (3) at (1.25, -5.5) {In-situ initialization};
		\node [style=greenLargeRec] (4) at (1.25, -7.75) {In-situ post-processing};
		\node [style=greenLargeRec] (5) at (1.25, -9) {In-situ finalization};
		\node [style=none] (7) at (3.25, -8.25) {};
		\node [style=none] (8) at (3.25, -6.25) {};
		\node [style=none] (9) at (1.25, -8.25) {};
		\node [style=none] (10) at (1.25, -6.25) {};
		\node [style=none] (11) at (4.25, -4) {};
		\node [style=none] (12) at (4.25, -10) {};
		\node [style=none] (18) at (1.25, -3.75) {(a) Application with synchronous in-situ task};
		\node [style=none] (19) at (9.25, -3.75) {(b) Application with asynchronous in-situ task};
		\node [style=none] (20) at (19, -3.75) {(c) Application with hybrid in-situ task};
		\node [style=none] (21) at (14, -4) {};
		\node [style=none] (22) at (14, -10) {};
		\node [style=blueLargeRec] (23) at (21.25, -4.75) {Application initialization};
		\node [style=blueLargeRec] (24) at (21.25, -6.75) {Application iteration};
		\node [style=blueLargeRec] (25) at (21.25, -9.75) {Application finalization};
		\node [style=greenLargeRec] (26) at (21.25, -5.5) {In-situ initialization};
		\node [style=greenLargeRec] (27) at (21.25, -7.5) {In-situ post-processing};
		\node [style=greenLargeRec] (28) at (21.25, -9) {In-situ finalization};
		\node [style=none] (30) at (23.25, -8.25) {};
		\node [style=none] (31) at (23.25, -6.25) {};
		\node [style=none] (32) at (21.25, -8.25) {};
		\node [style=none] (33) at (21.25, -6.25) {};
		\node [style=blueLargeRec] (34) at (11.5, -4.75) {Application initialization};
		\node [style=blueLargeRec] (35) at (11.5, -7.25) {Application iteration};
		\node [style=blueLargeRec] (36) at (11.5, -9.75) {Application finalization};
		\node [style=greenLargeRec] (37) at (11.5, -5.5) {In-situ initialization};
		\node [style=redLargeRec] (38) at (6.75, -7.25) {In-situ post-processing};
		\node [style=greenLargeRec] (39) at (11.5, -9) {In-situ finalization};
		\node [style=none] (41) at (13.5, -8.25) {};
		\node [style=none] (42) at (13.5, -6.25) {};
		\node [style=none] (43) at (11.5, -8.25) {};
		\node [style=none] (44) at (11.5, -6.25) {};
		\node [style=redLargeRec] (49) at (6.75, -4.75) {In-situ initialization};
		\node [style=redLargeRec] (50) at (6.75, -9.75) {In-situ finalization};
		\node [style=none] (51) at (4.75, -8.25) {};
		\node [style=none] (52) at (4.75, -6.25) {};
		\node [style=none] (53) at (6.75, -8.25) {};
		\node [style=none] (54) at (6.75, -6.25) {};
		\node [style=redLargeRec] (55) at (16.5, -7.5) {In-situ post-processing};
		\node [style=none] (58) at (18.25, -7.5) {};
		\node [style=none] (60) at (19.5, -7.5) {};
		\node [style=redLargeRec] (61) at (16.5, -4.75) {In-situ initialization};
		\node [style=redLargeRec] (62) at (16.5, -9.75) {In-situ finalization};
		\node [style=none] (63) at (14.5, -8.25) {};
		\node [style=none] (64) at (14.5, -6.25) {};
		\node [style=none] (65) at (16.5, -8.25) {};
		\node [style=none] (66) at (16.5, -6.25) {};
		\node [style=blueLargeRec] (69) at (4.25, -10.75) {Original application};
		\node [style=greenLargeRec] (70) at (8.5, -10.75) {Synchronous in-situ task};
		\node [style=redLargeRec] (71) at (13.25, -10.75) {Asynchronous in-situ task};
		\node [style=none] (72) at (17.25, -10.5) {ADIOS2 data transfer};
		\node [style=none] (73) at (16.5, -11) {};
		\node [style=none] (74) at (18, -11) {};
		\node [style=none] (75) at (8.5, -7.25) {};
		\node [style=none] (76) at (10, -7.25) {};
	\end{pgfonlayer}
	\begin{pgfonlayer}{edgelayer}
		\draw [style=myEdgeArrows] (0) to (3);
		\draw [style=myEdgeArrows] (1) to (4);
		\draw [style=myEdgeArrows] (4) to (5);
		\draw [style=myEdgeArrows] (5) to (2);
		\draw [style=myEdge] (9.center) to (7.center);
		\draw [style=myEdge] (7.center) to (8.center);
		\draw [style=myEdgeArrows] (8.center) to (10.center);
		\draw [style=myEdge] (11.center) to (12.center);
		\draw [style=myEdge] (21.center) to (22.center);
		\draw [style=myEdgeArrows] (23) to (26);
		\draw [style=myEdgeArrows] (24) to (27);
		\draw [style=myEdgeArrows] (27) to (28);
		\draw [style=myEdgeArrows] (28) to (25);
		\draw [style=myEdge] (32.center) to (30.center);
		\draw [style=myEdge] (30.center) to (31.center);
		\draw [style=myEdgeArrows] (31.center) to (33.center);
		\draw [style=myEdgeArrows] (34) to (37);
		\draw [style=myEdgeArrows] (39) to (36);
		\draw [style=myEdge] (43.center) to (41.center);
		\draw [style=myEdge] (41.center) to (42.center);
		\draw [style=myEdgeArrows] (42.center) to (44.center);
		\draw [style=myEdgeArrows] (37) to (35);
		\draw [style=myEdgeArrows] (35) to (39);
		\draw [style=myEdgeArrows] (38) to (50);
		\draw [style=myEdge] (53.center) to (51.center);
		\draw [style=myEdge] (51.center) to (52.center);
		\draw [style=myEdgeArrows] (52.center) to (54.center);
		\draw [style=dataflow] (60.center) to (58.center);
		\draw [style=myEdgeArrows] (55) to (62);
		\draw [style=myEdge] (65.center) to (63.center);
		\draw [style=myEdge] (63.center) to (64.center);
		\draw [style=myEdgeArrows] (64.center) to (66.center);
		\draw [style=dataflow] (73.center) to (74.center);
		\draw [style=myEdgeArrows] (3) to (1);
		\draw [style=dataflow] (76.center) to (75.center);
		\draw [style=myEdgeArrows] (49) to (38);
		\draw [style=myEdgeArrows] (61) to (55);
		\draw [style=myEdgeArrows] (26) to (24);
	\end{pgfonlayer}
\end{tikzpicture}
}%
\vspace{-0.5em}
\caption{Illustration of workflow applications with synchronous, asynchronous and hybrid in-situ tasks.}
\vspace{-1.5em}
\label{method_illustration}
\end{figure*}
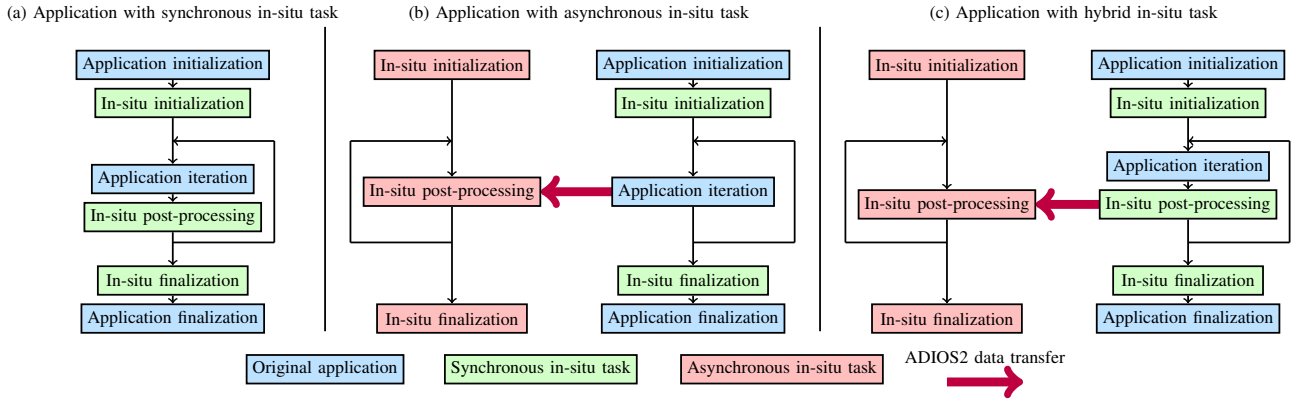

With the continuous development of central processing units (CPUs) and graphic processing units (GPUs), the peak performance of current High-Performance Computing (HPC) systems increases rapidly. This development has facilitated data-intensive research in various domains, enabling large-scale simulation. For instance, in the Molecular Dynamics (MD) domain, HPC systems are used to perform simulation and, from the simulated results, the physical movements of atoms and molecules can be described. Another example is the Computational Fluid Dynamics (CFD) domain, where computationally expensive numerical methods (executed on HPC systems) can help to analyze fluid flow problems. At the same time, due to limited walltimes, checkpointing is crucial for long runs on HPC systems. It also requires the storage of large amount of data. 
However, the input/output (IO) subsystem on HPC systems is developing relatively slowly, compared to the computational power, and the storage capacity is also limited. Traditionally, the results from one application would be stored via the IO system to the storage system.
The data analysis application reads these data back via the IO system from storage. Thus data has to go through the IO bottleneck twice. This problem may limit the efficiency of using the HPC system, the actual performance of the applications, which are originally designed to leverage the full computational capacity of the HPC resources, and, therefore,  scientific discovery. Our previous work~\cite{ju2022understanding} proved that, in-situ techniques~\cite{childs2020terminology} can avoid this problem for CPU-based applications, or at least reduce data traffic.
In-situ techniques can be classified into three types (Fig.~\ref{method_illustration}), depending on whether in-situ task(s) would interrupt the original application. In the \textit{synchronous approach} (Fig.~\ref{method_illustration}(a)), the original application would stop when the in-situ task is performed, and after the end of the in-situ task, it restarts. In the \textit{asynchronous approach}, (Fig.~\ref{method_illustration}(b)), before the execution, part of the computing resources are assigned to the in-situ tasks, and the original application would send the data to these separate resources, after which both the in-situ task and the original application continue concurrently. Lastly, in the \textit{hybrid approach} (Fig.~\ref{method_illustration}(c)), part of the in-situ task(s) would stop the original application, and the rest would be executed on separate computing resources. 

Current HPC systems often leverage the high peak performance of GPUs. When using GPUs, many applications are focusing on maximizing the efficiency of GPU usage, which brings the biggest performance benefit, but can result in the underutilization of CPUs on such GPU nodes.  
This makes these CPUs a perfect target for in-situ techniques, using them for data analysis, thereby increasing their utilization and, thus, the overall performance of the system, while the simulation mostly executes on the GPUs. 
By utilizing the CPUs on GPU nodes, in-situ tasks can significantly improve performance and optimize the use of system resources. However, in-situ tasks share HPC resources with the original application, which can introduce overhead and discourage developers from using in-situ techniques. It is essential to carefully evaluate the potential overhead and the trade-offs between reduced IO requirements and increased workloads before deploying in-situ methods. 

In our previous work, we studied the in-situ techniques applied on CPU-based applications.
This paper further expands the study to GPU-accelerated applications and compares the impact of in-situ approaches using three GPU use cases at scale. The following are the paper's specific contributions:
\begin{enumerate}
    \item it presents novel real-world case studies of applications in the CFD and MD fields. These case studies demonstrate the practicality of utilizing in-situ approaches for solving complex problems in these fields;
    \item it compares the impact of in-situ approaches on GPU-accelerated applications, providing valuable insights into the benefits and limitations of each approach;
    \item it analyzes critically which in-situ approach brings the least overhead to the original GPU-accelerated applications, highlighting the potential for in-situ methods to significantly improve the utilization of HPC resources and reduce IO overheads. 
\end{enumerate}

The rest of the paper is organized as follows: Section~\ref{relate_work} contains a summary of related work on in-situ techniques and case studies; Section~\ref{method} introduces the paper’s selected in-situ workflows and experimental setups; Section~\ref{case_study} contains information about the use cases and presents results and analyses; Section~\ref{conclusion} summarizes and discusses this paper.

\section{Related Work}
\label{relate_work}
In-situ visualization, as one of the common in-situ tasks, attracts interest from many researchers and developers. VisIt with Libsim~\cite{childs2012visit,kuhlen2011parallel}, ParaView with Catalyst~\cite{ayachit2015paraview} and SENSEI~\cite{ayachit2016sensei} are common in-situ visualization systems based on the Visualization Toolkit (VTK) data format~\cite{schroeder1998visualization}. However,  VTK is difficult to use when dealing with tasks other than visualization, and hence we decided not to rely on VTK for the non-visualization in-situ tasks. 
\textit{Dorier et al.}~\cite{dorier2022colza,dorier2023towards} developed a data staging service, \textit{Colza}, to enable elastic in-situ visualization with HPC simulations. 

Compared to these approaches, this paper focuses also on other (non-visualization) in-situ tasks. The Adaptable IO System (ADIOS)~\cite{godoy2020adios, liu2014hello}, initially intended as a higher-level IO abstraction, can also be utilized for in-situ processing. Its independence from VTK and compatibility with various data formats make it an ideal choice for a versatile in-situ framework, which we are aiming at. 
\textit{Gainaru et al.}~\cite{gainaru2022understanding} studied the impact of data staging in asynchronous in-situ techniques to couple simulation and in-situ analyses. 

Although in-situ data analysis has become increasingly popular in scientific computing, only a limited number of researches have been studying how to exploit heterogeneous CPU/GPU systems for in-situ analysis.  \textit{Xing et al.}~\cite{xing2022gpu} proposed a performance modelling methodology that aims to predict the optimal placement (CPU or GPU) and data representation choice (summarization or original) for a given configuration. \textit{Hagan et al.}~\cite{hagan2011multi} demonstrated the effectiveness of a load balancing method using an N-body simulation and a ray tracing visualization with varying input size, supersampling, and simulation parameters. \textit{Qin et al.}~\cite{qin2022gpu} focus on optimizing visualization methods by directly accessing DEM simulation results, exchanging data between GPUs, and accelerating pixel composition using GPUs.

\textit{Mittal and Vetter}~\cite{mittal2015survey} provided a survey of techniques to fully use the heterogeneous HPC systems. They also addressed resource underutilization and the challenge of balancing the workload between GPUs and CPUs. However, they focused on the techniques for a single application to use heterogeneous HPC systems. This paper studies how to couple applications and in-situ tasks without redesigning code structure. 

\section{Methodology}
\label{method}

This section presents our synchronous, asynchronous, and hybrid in-situ workflow and introduces the experimental setup.

\subsection{In-situ workflow}

The first challenge faced by integrating in-situ tasks is to compile and link the original codes with the in-situ tasks, which can often be problematic because they may be developed using different programming languages. To overcome this issue, we incorporate adaptor functions into our workflow design. 
We use the ADIOS2 framework to transfer data between the original application and in-situ tasks in both asynchronous and hybrid in-situ approaches. Because ADIOS2 provides Fortran, C/C++, and Python APIs, it eases the development of adaptor functions.

In the synchronous approach, data are passed from the original application to the in-situ processing (Fig.~\ref{method_illustration}(a)). 
If the in-situ task uses a different data structure, the adaptor functions may need to perform a deep copy. 
Since the original application is halted during the in-situ execution, 
data consistency is guaranteed, and 
no data transfer using the ADIOS2 library is necessary. However, when the original application uses GPUs for acceleration and the in-situ task is CPU-only, additional data synchronization between GPU and CPU is required.

In the asynchronous approach, the data is transmitted to the in-situ task via a writer and reader pair using the "insituMPI" engine from ADIOS2 (Fig.~\ref{method_illustration}(b)), which is based on MPI communication. The original application and in-situ task are launched concurrently in a multiple-program multiple-data (MPMD) mode. Workloads are distributed across separate computational resources. The total number of available resources, $p_{t}$, can be assigned in various chunks to the original application $p_{o}$ and in-situ task $p_{i}$ such that $p_{o} + p_{i} = p_{t}$. 
To ensure data consistency, the original application needs to wait for the end of the MPI communication. After receiving the data from the ADIOS2 reader, the in-situ task is executed concurrently with the original application. 
If the original application and in-situ task have different structures, the adaptor functions can also perform the necessary adaptations. 
This approach has some small but unavoidable overhead, including the communication between the original application and in-situ task, the no-overlapping execution before the first data traffic between the original application and in-situ task and the last execution of the in-situ task, which is not overlapping with the original application.
Data synchronization between GPU and CPU is necessary for the GPU-accelerated original application if ADIOS2 is not built with CUDA-aware MPI.

The hybrid in-situ approach (Fig.~\ref{method_illustration}(c)) consists of both synchronous and asynchronous components. Adaptor functions, as in the synchronous approach, pass data to the synchronous part of the in-situ task. After this, intermediate data are sent to the asynchronous part of the in-situ task via ADIOS2, as in the asynchronous approach. 
In this approach, the original application is directly linked and compiled with the synchronous part of the in-situ task, and it is launched concurrently in MPMD mode with the asynchronous part of the in-situ task.



\subsection{Experimental setup}
\label{exp_setup}
We use the Raven supercomputer at the Max Planck Computing and Data Facility (MPCDF).~\cite{Raven} One Raven CPU node contains two Intel Xeon IceLake-SP 8360Y processors with 36 cores each and 256 GB RAM. In addition, Raven also provides GPU-accelerated compute nodes, each with 4 Nvidia A100 GPUs (4 × 40 GB HBM2 memory per node and NVLink) and two Intel Xeon CPUs with 512 GB RAM. We use MPMD configuration files to define how CPU cores are allocated to the original application and in-situ tasks for the asynchronous and hybrid in-situ approach. To ensure the data transfer within one node, we dedicate one set of CPU cores to the original application, while the rest are dedicated to the in-situ task. For the GPU-accelerated codes, we use Nvidia Multi-Process Service (MPS)~\cite{noauthor_introduction_nodate} to allow multiple CPU cores to access the same GPU and to use the GPUs more efficiently and allocate cores for the original application and in-situ tasks evenly on two CPUs because, on one Raven GPU node, one CPU is directly connected to two GPUs. 

We profile the CPU and GPU usages with the MPCDF HPC monitoring system~\cite{stanisic2020mpcdf} and NVIDIA's NSight Systems~\cite{noauthor_nsight_nodate}. The HPC performance monitoring system is extremely lightweight and runs in the background, invisibly to the users, while it provides sophisticated performance reports. The NSight Systems provides detailed information about GPU utilization, including the timelines including CUDA kernels and memory operation and the size of memory traffic. 

\begin{figure}[t]
\centering
	\vspace{-1.em}
	\includegraphics[width=0.45\textwidth]{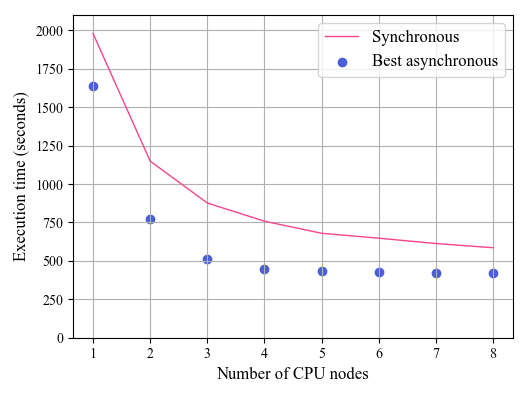}
	\vspace{-1.5em}
	\caption{Execution time of CPU-based NEKO with synchronous and asynchronous image generation on various numbers of fully used Raven CPU node(s)}
	\vspace{-1.5em}
	\label{neko_img_cpu}
\end{figure}

\section{Case Studies}
\label{case_study}
We first discuss a CFD use-case (turbulence simulations) followed by a Molecular Dynamics use case. For the turbulence case study, we use the incompressible spectral-element Navier-Stokes solver, NEKO~\cite{jansson2021neko}, and two common in-situ tasks to investigate the impact of in-situ tasks on GPU-based CFD simulations: data visualization and lossy and lossless compression; for the MD simulation case study, we use the density function theory solver, Quantum-Espresso (QE)~\cite{giannozzi2009quantum}, and lossless compression as in-situ task.



\subsection{Turbulence with image generation} 

The spectral element method (SEM)~\cite{deville2002high} is a high-order Finite element method used in state-of-the-art, high-fidelity CFD simulations because of its many good properties in regards to accuracy \cite{rezaeiravesh2021numerical}. From the computational standpoint some of the advantages of SEM are the possibility to implement it in a matrix-free fashion, avoiding the explicit construction of any operator matrix, and its weak element coupling which allows operations to be mostly performed on a local basis, reducing communication requirements. These characteristics, among others, allow the method to handle large problems and perform efficiently on large number of processing elements~\cite{karp2022reducing}.
NEKO~\cite{jansson2021neko} is a portable framework that implements SEM in object oriented modern Fortran, allowing better control on memory allocation and modularity and thus providing support for multiple compute architectures.  

The first in-situ task integrated into NEKO is data visualization, usually needed to analyze the features of the fluid flow on a time dependent case. We use ParaView with Catalyst as the image generator, with a workflow that generates a VTK grid during initialization and reads a customized ParaView Pipeline Python script. This Python script defines how the ParaView with Catalyst coprocessor renders the output image, including details such as camera position, image size, and slice position. 
To compare the different impacts of in-situ techniques on CPU-based and GPU-accelerated NEKO, we first studied the CPU-based NEKO with in-situ image generation. In the synchronous approach, the visualization is executed on the same cores as the NEKO. In the asynchronous approach, the visualization is executed on cores different from the NEKO simulation, but still on the same node. 

We used the three-dimensional Taylor-Green Vortex (TGV) simulation as our benchmark case. For the CPU-based NEKO, we used a mesh with $32^3$ elements and ran 2000 simulation steps with image generation every 20 simulation steps. We ran strong scalability tests to analyze the performance of synchronous and asynchronous approaches. For the synchronous approach, we tested the performance of 1 to 8 CPU node(s) using all 72 cores on each node. For the asynchronous approach, to study the influence of resource allocation, we assigned  2, 4, 8, 12, 18, 24, and 36 cores out of 72 cores on each node for the asynchronous image generation.

\begin{table}[t]
\vspace{-2em}
\centering
\caption{Configurations of the asynchronous image generation with the best performance}
\vspace{-1em}
\begin{tabular}{|c|cc|}
\hline
\textbf{\begin{tabular}[c]{@{}c@{}}Number of \\ CPU nodes\end{tabular}} & \textbf{\begin{tabular}[c]{@{}c@{}}Number of CPU cores\\ per node for NEKO\end{tabular}} & \textbf{\begin{tabular}[c]{@{}c@{}}Number of CPU cores per\\ node for image generation\end{tabular}} \\ \hline
1                                                                       & 70                                                                                       & 2                                                                                                    \\
2                                                                       & 70                                                                                       & 2                                                                                                    \\
3                                                                       & 68                                                                                       & 4                                                                                                    \\
4                                                                       & 60                                                                                       & 12                                                                                                   \\
5                                                                       & 60                                                                                       & 12                                                                                                   \\
6                                                                       & 54                                                                                       & 18                                                                                                   \\
7                                                                       & 48                                                                                       & 24                                                                                                   \\
8                                                                       & 48                                                                                       & 24                                                                                                   \\ \hline
\end{tabular}
	\vspace{-2.em}
\label{neko_img_config}
\end{table}


We first verified that our synchronous and asynchronous in-situ image generation could generate the same images as the traditional visualization from the simulation result. 
The image generated in the in-situ task is identical to the one from the traditional post-processing visualization from the saved simulation result. At the same time, we avoided an 8 GB VTK file for each step with the in-situ techniques.

Like the experimental results in our previous work to integrate the in-situ image generation into the CFD solver Nek5000~\cite{ju2022understanding}, the asynchronous image generation can outperform the synchronous one, and the best performance of the asynchronous approach appears when the simulation and image generation take about the same amount of time. In that case, the computing resources are allocated optimally according to the workload. As Fig.~\ref{neko_img_cpu} shows, by selecting an appropriate number of ranks for both NEKO and in-situ image generation, the asynchronous approach yields shorter total execution times than the synchronous approach. However, the optimal number of resources assigned to the in-situ task varies based on the number of nodes used (TABLE~\ref{neko_img_config}). The more CPU nodes are used, the more CPU cores are required for in-situ image generation because of the worse scalability of the image generation compared to the scalability of  NEKO. This leads to a change in the workload ratio between simulation and in-situ task and influences the resource allocation. 

\begin{figure}[t]
\centering
	\vspace{-2.em}
	\includegraphics[width=0.45\textwidth]{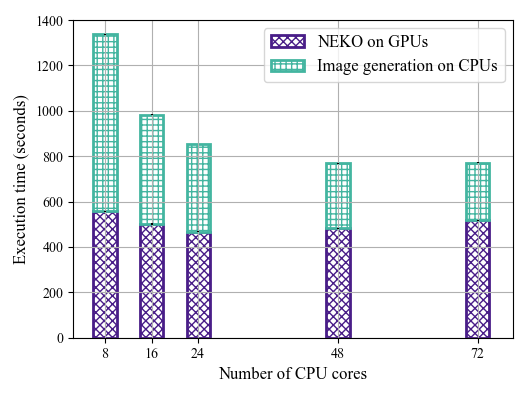}
	\vspace{-1.5em}
	\caption{Execution time of GPU-accelerated NEKO with synchronous image generation on two Raven GPU nodes with full usage of eight GPUs and various numbers of CPU cores}
	\vspace{-1.5em}
	\label{neko_img_gpu_sync}
\end{figure}

\begin{figure*}[t]
	\vspace{-2em}
\centering
	\includegraphics[width=0.95\textwidth]{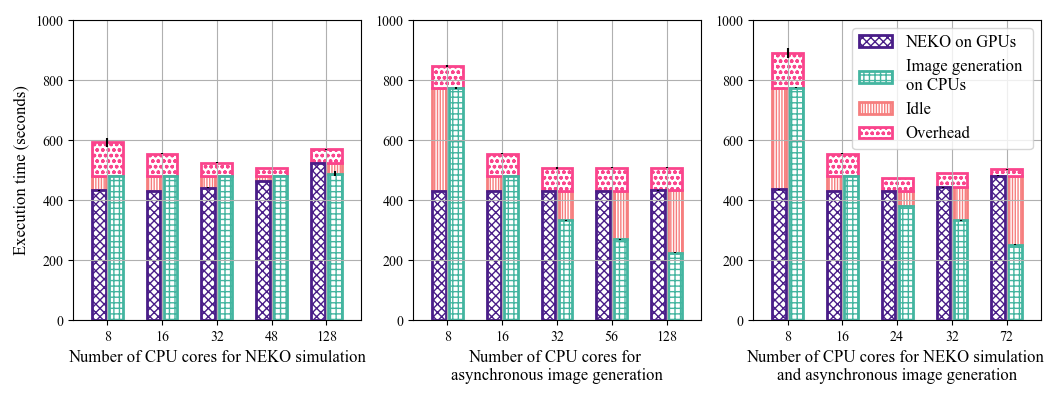}
	\vspace{-1.5em}
	\caption{Execution time of GPU-accelerated NEKO with asynchronous image generation every 50 simulation steps on two Raven GPU nodes with various CPU cores for NEKO and 16 CPU cores for image generation (left), 16 CPU cores for NEKO and various CPU cores for image generation (middle) and the same number of CPU cores for NEKO and image generation (right). In all cases, all eight GPUs on the GPU nodes are used for NEKO}
	\vspace{-1.5em}
	\label{neko_img_gpu_async}
\end{figure*}

\begin{figure}[t]
\centering
	\includegraphics[width=0.45\textwidth]{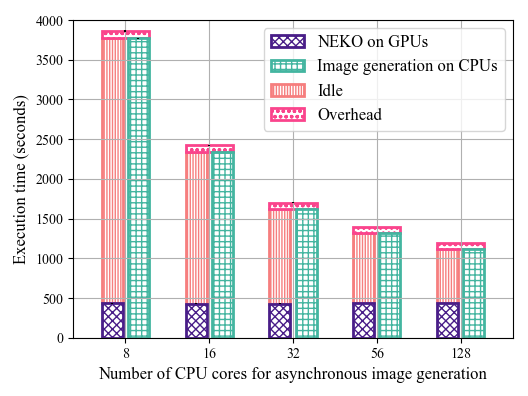}
	\vspace{-1.5em}
	\caption{Execution time of GPU-accelerated NEKO with asynchronous image generation every ten simulation steps on two Raven GPU nodes with 16 CPU cores for NEKO and various CPU cores for image generation. All eight GPUs on the GPU nodes are used for NEKO}
	\vspace{-1.5em}
	\label{neko_img_gpu_async2}
\end{figure}

\begin{figure*}[t]
\centering
	\vspace{-2em}
	\includegraphics[width=0.95\textwidth]{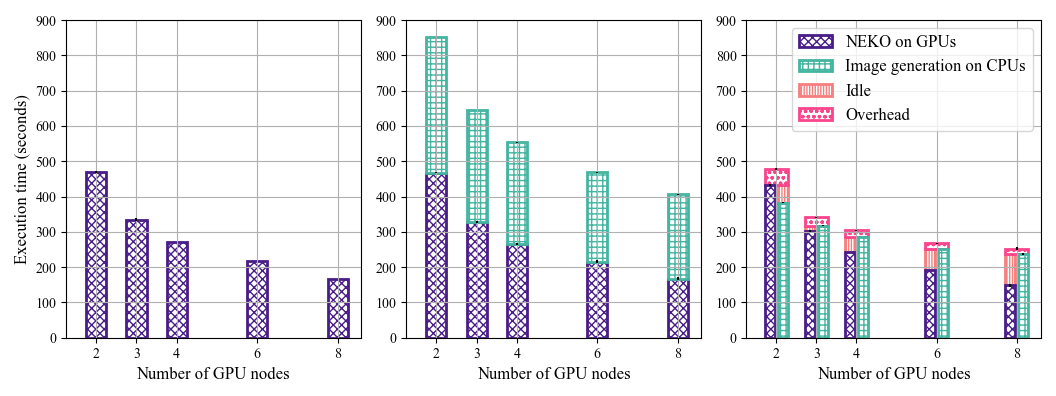}
	\vspace{-1.5em}
	\caption{Execution time of GPU-accelerated NEKO without in-situ tasks using 12 CPU cores per node (left) and GPU-accelerated NEKO with synchronous image generation using 12 CPU cores per node (middle) and asynchronous image generation using 12 CPU cores per node and another 12 CPU cores for NEKO (right) on a various number of Raven GPU nodes. All four GPUs on each GPU node are used for NEKO. }
	\vspace{-1.5em}
	\label{neko_img_gpu}
\end{figure*}

Then we studied the GPU-accelerated NEKO with in-situ visualization. In the synchronous approach, the simulation is executed mainly on GPUs, while the visualization is executed on CPU cores. However, the GPUs still wait until the visualization on CPU cores is done because those CPU cores are also responsible for parts of the NEKO simulation, including transferring the data between the GPUs and CPUs and launching GPU kernels. In the asynchronous approach, in contrast, the visualization is executed on  CPU cores separate from the ones used for NEKO, and the GPUs only wait until the data is sent. For this test, we used a mesh with $64^3$ elements and ran 2000 simulation steps. The mesh size was increased compared to the CPU case to efficiently use the significantly higher computational capacity of the GPUs.  We generated images every 50 simulation steps, except when we studied the influence of the in-situ task workload on the total execution. 
We used two GPU nodes with all eight GPUs on these nodes used for NEKO and 4, 8, 12, 24, and 36 CPU cores on each GPU node to study the influence of used CPU cores number on the performance of the NEKO with synchronous image generation. In the asynchronous approach, the image generation was executed by additional CPU cores on the same GPU node. We first designed two groups of experiments on two Raven GPU nodes: in the first group of experiments, we used 16 CPU cores for asynchronous image generation and used 8, 16, 32, 48, and 128 CPU cores for NEKO; in the second group, we used 16 CPU cores for NEKO and used 8, 16, 32, 48, and 128 CPU cores to generate images asynchronously; in the third group, we used the same number of CPU cores for NEKO and asynchronous image generation respectively, i.e. 8, 16, 24, 32, and 72 CPU cores. 
Then we tested the influence of the in-situ task workload. We repeated the second group experiments with image generation every ten simulation steps. 
We avoid a 26 GB VTK file per in-situ step for larger case on GPU nodes. 

In Fig.~\ref{neko_img_gpu_sync}, the total execution time of the synchronous approach on two nodes decreases when the number of CPU cores increases because of the reduced time for image generation with increased resources, while NEKO's execution stays mostly the same using the same number of GPUs in all cases.  

The left plot in Fig.~\ref{neko_img_gpu_async} shows how the number of CPU cores for NEKO influences the total execution time when we fixed the number of CPU cores used for asynchronous image generation. The performance difference in this group of experiments is small because the time to generate image asynchronous stays constant when the same amount of CPU cores are used, and the constant number of GPUs used leads to the NEKO time only fluctuating slightly. The middle plot in Fig.~\ref{neko_img_gpu_async} shows that when the number of CPU cores for NEKO stays the same, the total execution time first decreases with the increasing number of CPU cores for asynchronous image generation until the image generation takes the same amount of time as the NEKO and then stays the same. Because the total execution time depends on the longer execution of simulation or image generation.
The right plot in Fig.~\ref{neko_img_gpu_async} shows how the performance changes when the numbers of CPU cores for NEKO and for image generation are the same. The total execution time first decreases with the increasing number of CPU cores for NEKO and image generation until 24 cores (12 cores on each node) are used for NEKO, and another 24 cores are used for image generation because of the decreasing image generation time. Then the total execution time increases slightly because of the slight increment in NEKO time.

When the frequency of generating the image increases, the workload of the in-situ task compared to the simulation increases. As shown in Fig.~\ref{neko_img_gpu_async2}, although the total execution time of the GPU-accelerated NEKO with asynchronous image generation every ten simulation steps decreases with the increasing number of CPU cores for image generation, even when all the originally idle CPU cores are used for image generation, i.e. 128 additional CPU cores, the asynchronous image generation takes longer than the NEKO.

\begin{figure}[t]
\centering
	\includegraphics[width=0.45\textwidth]{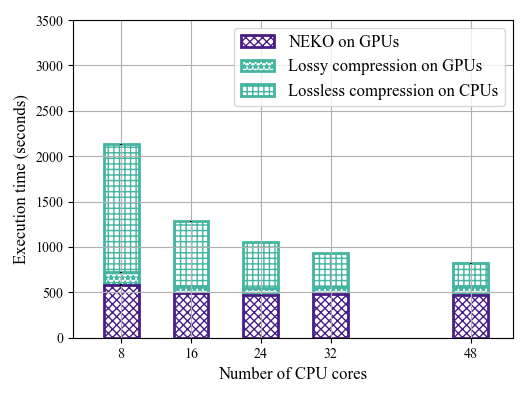}
	\vspace{-1.5em}
	\caption{Execution time of GPU-accelerated NEKO with synchronous lossy and lossless compression on two Rave GPU nodes, with all eight GPUs and various numbers of CPU cores used.}
	\vspace{-1.5em}
	\label{neko_comp_gpu_sync}
\end{figure}

All the experiments above are executed on two Raven GPU nodes. It is the lowest possible number of nodes to simulate this TGV case on GPUs because of memory limitations. To further understand the influence of the in-situ technique, we compared the performance of GPU-accelerated NEKO with synchronous and asynchronous image generation every 50 simulation steps with the original NEKO on multiple Raven GPU nodes, specifically on 2, 3, 4, 6, and 8 nodes. In all the experiments, we used all four GPUs on each node. In the original NEKO and NEKO with synchronous image generation, we used 12 CPU cores on each node; in the asynchronous approach, we used 12 CPU cores on each node for NEKO and another 12 CPU cores on each node for image generation, which is the configuration of the fastest asynchronous approach on two Raven nodes. As shown in Fig.~\ref{neko_img_gpu}, the NEKO time stays almost the same regardless of the in-situ approach. The middle plot in Fig.~\ref{neko_img_gpu} shows that the time to generate the images synchronously barely decreases when more nodes are used because of the collective communication overhead and consequent poor scalability in image generation. In contrast, the right plot in Fig.~\ref{neko_img_gpu} shows that the difference between NEKO without in-situ tasks and NEKO with asynchronous image generation increases with the number of nodes used. Only a small communication overhead is added to the NEKO when two or three Raven GPU nodes are used. In these cases, the asynchronous image generation takes shorter than or almost the same as the NEKO. Because of the image generation's worse scalability, the total execution time depends on how long asynchronous image generation takes when four to eight nodes are used. In general, the asynchronous approach outperforms the synchronous approach.

We studied GPU usage with the MPCDF HPC monitoring system. With 24 CPU cores on two Raven GPU nodes, we achieved $58.5987\%$ average GPU usage and $82.117\%$ average GPU memory usage; with 72 CPU cores, these numbers increase to $64.583\%$ and $99.881\%$ respectively. 
Thanks to MPS, when more CPU cores are used for NEKO, the GPU usage on two Raven GPU nodes can be slightly improved. And the GPU usage does not change when the in-situ task is integrated into  NEKO. But the timelines in the NSight profiling results show that the synchronous approach brings stops in GPUs, while, in the asynchronous approach, GPUs have no visible stop compared to the original NEKO. The extra data synchronization between GPUs and CPUs in both in-situ approaches brings additional data transfer from the GPU to one CPU core.
But the additional time of this memory operation is smaller than the performance fluctuation of original NEKO. 

\begin{figure}[t]
\centering
 	\includegraphics[width=0.45\textwidth]{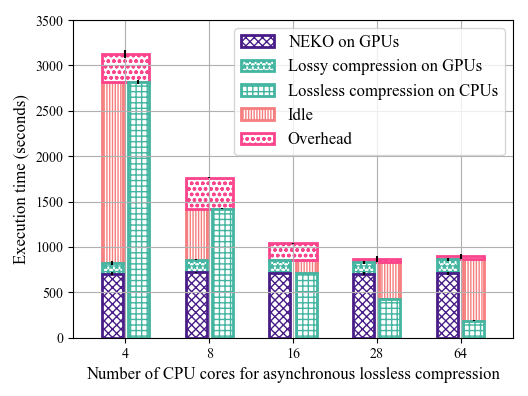}
	\vspace{-1.5em}
	\caption{Execution time of GPU-accelerated NEKO with hybrid lossy and lossless compression on two Raven supercomputer nodes with all eight GPUs and eight CPU cores for NEKO with synchronous lossy compression and a various number of CPU cores for asynchronous lossless compression.}
	\label{neko_comp_gpu_hybrid}
	\vspace{-2em}
\end{figure}

\subsection{Turbulence with data compression} 

\begin{figure*}[t]
\centering
	\vspace{-2em}
	\includegraphics[width=0.95\textwidth]{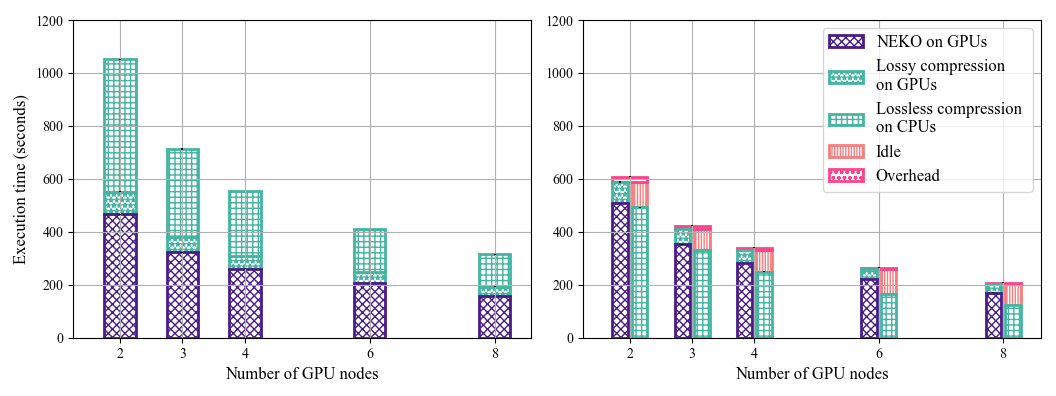}
	\vspace{-1.5em}
	\caption{Execution time of GPU-accelerated NEKO with synchronous data compression using 12 CPU cores pre node (left) and asynchronous lossless compression using 12 CPU cores per node and another 12 CPU cores for NEKO with synchronous lossy compression (right) on various numbers of Raven GPU nodes. All four GPUs on each GPU node are used for NEKO. }
	\vspace{-1.5em}
	\label{neko_comp_gpu}
\end{figure*}

The lossy and lossless data compression is the second in-situ task integrated into NEKO. Lossy compression is a physics-based method inspired by JPEG compression standard~\cite{wallace1992jpeg}. Turbulence is a complex phenomenon characterized by chaotic motion at multiple scales, with only a small subset of motions containing the majority of energy in the flow. \textit{Otero et al.}~\cite{otero2018lossy} proposed a technique that allows the retention of only the data associated with the most energetic motions while discarding the remaining data. We reuse the functions from NEKO, so this part of the in-situ task is one example of a deep coupled in-situ task. For the lossless compression, we use the embedded Bzip2 lossless compression functions in the ADIOS2 library in the synchronous approach. In this case, we test the synchronous and hybrid in-situ approaches. 


The amount of data physics-based lossy compression together with the lossless compression can compress depends on the maximal allowed error. With it equal to $10^{-2}$, we can compress about 98\% data while keeping sufficient accuracy.~\cite{ju2022understanding}


In the synchronous approach, the lossy compression is executed on the same GPUs as NEKO, and the lossless compression is executed on the CPU; in a hybrid approach, the lossy compression is executed on the GPU synchronously, and the lossless compression is performed on separate CPU cores asynchronously. We used the mesh with $64^3$ elements.
We used two GPU nodes with all eight GPUs on these nodes used for NEKO and 4, 8, 12, 16, 20, and 24 CPU cores to study the influence of used CPU cores on the performance. In the hybrid approach, the lossy compression was ported to the same GPUs executing NEKO, while the lossless compression was executed by additional CPU cores on the same GPU node. We reported here one group of experiments on two Raven GPU nodes: we used eight GPUs and eight CPU cores for NEKO with synchronous lossy compression and used additional 4, 8, 16, 28, and 64 CPU cores to compress the data losslessly. 

The total execution time of the synchronous approach consists of NEKO time and data compression time (Fig.~\ref{neko_comp_gpu_sync}). It decreases with the increasing number of used CPU cores because the lossless compression time decreases.

As shown in Fig.~\ref{neko_comp_gpu_hybrid}, the time to perform asynchronous lossless  data compression decreases with the increasing number of additional CPU cores. The total execution time of the hybrid approach decreases with the time for asynchronous compression until the time for NEKO and synchronous lossy compression takes longer because the total execution time depends on the longer time for asynchronous lossless compression or the sum of the NEKO and synchronous lossy compression. The latter takes constant time because it is done on the same number of GPUs.

Fig.~\ref{neko_comp_gpu} shows our comparison of GPU-accelerated NEKO with synchronous and hybrid data compression every ten simulation steps on multiple Raven GPU nodes, i.e., on 2, 3, 4, 6, and 8 nodes. In all the experiments, we used all four GPUs on each node. In NEKO with synchronous data compression, we used 12 CPU cores on each node; in the hybrid approach, we used 12 CPU cores on each node for NEKO with lossy compression and another 12 CPU cores on each node for lossless compression, as it is also the configuration with the best performance. In general, both the synchronous and hybrid approach scale well with the number of GPU nodes used, and the hybrid approach outperforms the synchronous approach because the overhead of the hybrid approach is smaller than the execution time of lossless data compression in the synchronous approach. 
The additional time required by lossy data compression on GPUs is the dominant time increment in the hybrid approach. According to the NSight Systems profiling result, the most time-consuming kernels in lossy compression are two sorting kernels, which are less preferable on GPUs compared to computation. 

\subsection{Molecular Dynamics simulation with data compression} 



\begin{figure*}[t]
\centering
	\vspace{-2.em}
	\includegraphics[width=0.95\textwidth]{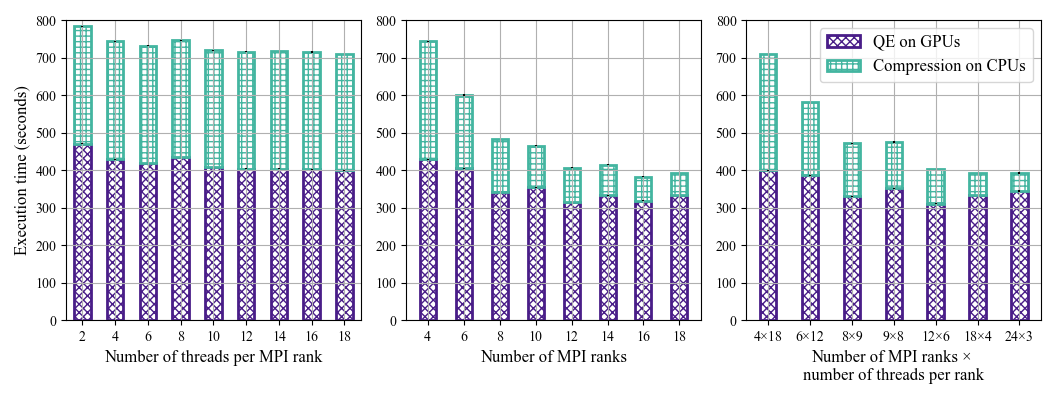}
	\vspace{-1.5em}
	\caption{Execution time of GPU-accelerated QE with synchronous lossless compression on one Raven GPU node with four MPI ranks (left), with four threads per MPI rank (middle) and with full usage of the node (right).}
	\vspace{-1.5em}
	\label{qe_comp_gpu_sync}
\end{figure*}

\begin{figure*}[t]
\centering
	\includegraphics[width=0.9\textwidth]{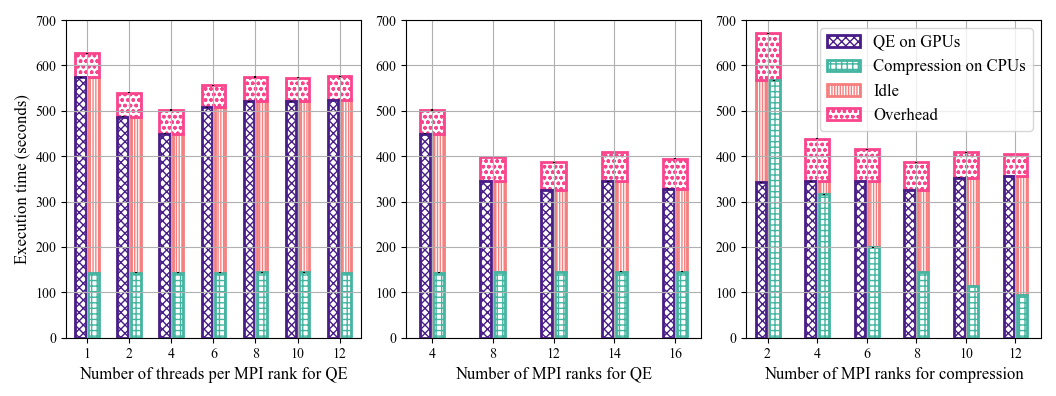}
	\vspace{-1.5em}
	\caption{Execution time of GPU-accelerated QE with asynchronous lossless compression on one Raven GPU node with four MPI ranks for QE and eight MPI ranks per node for compression (left), with four threads per MPI rank for QE and eight MPI ranks for compression (middle) and with 12 MPI ranks with four threads per MPI rank for QE (right).}
	\vspace{-1.5em}
	\label{qe_comp_gpu_async}
\end{figure*}

\begin{figure*}[t]
	\vspace{-2.em}
	\centering
	\includegraphics[width=0.95\textwidth]{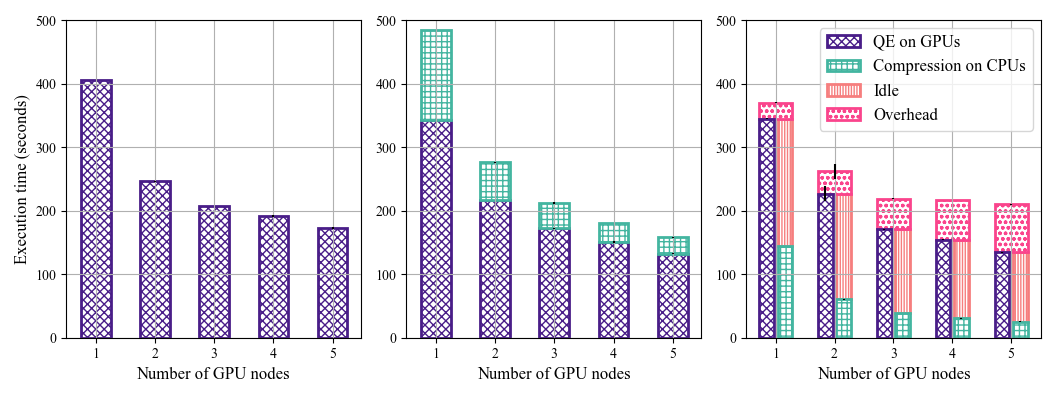}
	\vspace{-1.5em}
	\caption{Execution time of GPU-accelerated QE without compression using eight MPI ranks per node with four threads per MPI rank (left), with synchronous lossless compression using eight MPI ranks per node with four threads per MPI rank (middle) and with asynchronous lossless compression using eight MPI ranks per node with four threads per MPI rank for QE and eight MPI ranks per node for compression (right) on a various number of Raven GPU node(s).}
	\vspace{-1.5em}
	\label{qe_comp_gpu_sync_async}
\end{figure*}

 \begin{table}[t]
\caption{Compression ratio of different compression algorithms}
\vspace{-1.em}
\centering
\begin{tabular}{|c|c|}
\hline
\textbf{Compression algorithm} & \textbf{Compression ratio ($CR$)} \\ \hline
Bzip2~\cite{seward1996bzip2}                  & 1.5639\%          \\
LZ4~\cite{ziv1977universal}              & 4.5662\%          \\
LZ4HC                 & 5.7088\%          \\
ZLIB~\cite{gailly2004zlib}                  & 10.1870\%         \\
ZSTD~\cite{noauthor_facebookzstd_nodate}                  & 5.9271\%         \\
\hline
\end{tabular}
\vspace{-2.5em}
\label{compression_ratio}
\end{table}

Car-Parrinello (CP) molecular dynamics simulations~\cite{car1985unified} enable the characterization of complex electronic interactions in large molecules and small proteins accurately, making them valuable tools for improving drug design and understanding metabolic processes. These simulations, which use ab-initio density functional theory, require substantial computational resources, often scaling to hundreds or thousands of nodes and running for months with frequent checkpoints and restarts to accommodate workload management policies and system failures. 
Quantum-Espresso (QE)~\cite{giannozzi2009quantum} can conduct these simulations and supports OpenMP and MPI and offers OpenACC with CUDAFortran support for optimal performance on GPUs.

In CP simulation, one type of large restart file is the file storing the wave function coefficients. This file could be hundreds of gigabytes large in a large simulation case like the Coronavirus protein~\cite{noauthor_quantum_espressocp5r7y-covid19_nodate} and worth of compression. 

We chose the 50-step simulation of 256 water molecules as the benchmark case of GPU-accelerated QE, and compressed the wave function coefficients with ZLIB lossless compression on CPU every ten simulation steps because it has the highest compression ratio among the lossless compression algorithms we tested, as shown in TABLE~\ref{compression_ratio}. The compression ratio ($CR$) is calculated with the following equation: 
\begin{equation}
    CR = \frac{original\ size - compressed\ size}{original\ size}
\end{equation}

For the GPU-accelerated QE with synchronous and asynchronous compression, we first measured the execution time with different configurations on one Raven GPU node. All four GPUs on each node are used for QE in these tests. In the synchronous approach, we measured the execution time when 4 MPI ranks with 2, 4, 6, 8, 10, 12, 14, 16, and 18 threads per MPI rank were used, the execution time when 4, 6, 8, 10, 12, 14, 16, and 18 MPI ranks with 4 threads per MPI rank were used, and the execution time when all CPU cores on the GPU node were used, i.e., 4, 6, 8, 9, 12, 18 and 24 MPI ranks with 18, 12, 9, 8, 4 and 3 threads per MPI rank respectively. In the asynchronous approach, we tested the execution time when 4 MPI ranks with 2, 4, 6, 8, 10, and 12 threads per MPI rank for QE and 8 MPI ranks for compression were used, the execution time when 4, 8, 12, and 16 MPI ranks with 4 threads per MPI rank for QE and 8 MPI ranks for compression, and the execution time when 12 MPI ranks with 4 threads per MPI rank for QE and 2, 4, 6, 8, 10 and 12 MPI ranks for compression were used. We pinned one thread corresponding to a single CPU core on the GPU node in these experiments.

The number of threads per MPI rank used barely influences the total execution time (left plot in Fig.~\ref{qe_comp_gpu_sync}), because the compression library does not support multithreading and the number of threads per rank has a low impact on the QE time. Indeed, most of the OpenMP thread support for CPU-based QE is replaced in GPU-accelerated version with OpenACC to offload the workload on GPUs. 
According to the fact that compression is always done by one thread per MPI, the compression time decreases only when the number of MPI ranks is increased (middle plot in Fig.~\ref{qe_comp_gpu_sync}). Because most of the QE simulation has been ported to GPU, the QE simulation time decreases slightly until eight MPI ranks are used and then fluctuates with the increasing number of MPI ranks used. The total execution time is the sum of the simulation and compression time and decreases until 12 MPI ranks are used and then fluctuates because the time reduced in compression is smaller than the fluctuation in the QE simulation. A similar phenomenon also appears when the full GPU node is used (right plot in Fig.~\ref{qe_comp_gpu_sync}). Moreover, the best performance appears when 16 MPI ranks with four threads per MPI rank are used, although not all CPU cores on the GPU node are used. 

The left plot in Fig.~\ref{qe_comp_gpu_async} shows the execution time when four MPI ranks for QE and eight MPI ranks for compression are used. The QE simulation time and the total time first decreased with the increasing number of threads per MPI rank, until four threads per MPI rank for QE are used. Then the time increased slightly and stabilized. Similarly, the QE simulation time and thus the total time first decreased with the increasing number of MPI ranks for QE and then fluctuated as shown in the middle plot in Fig.~\ref{qe_comp_gpu_async}. The right plot in Fig.~\ref{qe_comp_gpu_async} shows that the compression time decreases with the increasing number of MPI ranks for compression, and the total time decreases until the QE simulation takes longer than the asynchronous compression. Generally, when 12 MPI ranks with four threads per MPI rank for QE and eight MPI ranks for compression are used, the GPU-accelerated QE with asynchronous compression performs best on one Raven GPU node. 

We also compared original GPU-accelerated QE, QE with synchronous compression and QE with asynchronous compression on one to five Rave GPU node(s) as shown in Fig.~\ref{qe_comp_gpu_sync_async}
In the original QE and the synchronous approach, we used eight MPI ranks per node with four threads per MPI rank; in the asynchronous approach, we used eight MPI ranks per node with four threads per MPI rank for QE and eight MPI ranks per node for compression because these are the configurations with the best performance in previous experiments. In both approaches, the total time decreases with increasing nodes. When one node is used, the asynchronous approach outperforms the synchronous approach, and more nodes are used, and the synchronous approach outperforms the asynchronous approach. Because when more nodes are used, the number of used CPU cores also increases, and compression is no longer computationally expensive for these CPU cores. At the same time, the communication overhead in the asynchronous approach increases. The original QE takes longer because the original QE used one MPI ranks to collect all necessary information and to write the checkpointing file. In our in-situ approaches, we replace this with new data structure and parallel IO or communication with ADIOS2 library. 
The scaling on GPUs of QE is generally limited by communications when using the default level of parallelism, which is used in this case. So usually, it does not scale much above the minimum number of GPU(s) required for memory, where the asynchronous approach is preferred.



\section{Discussion and Conclusions}
\label{conclusion} 
In this paper, we studied the impact of synchronous, asynchronous, and hybrid in-situ techniques on data-intensive GPU-accelerated simulations. When applications utilize  GPUs to accelerate their execution, it is more complicated to integrate in-situ task due to the extra library required by the GPU operation, such as CUDA and OpenACC. However, in typical GPU-accelerated applications most workload is executed on GPUs, leaving the associated CPU resources mostly unused. As shown in this paper, these underused CPU resources can be efficiently used for executing concurrent in-situ computations. 

Clearly, when using CPU cores for the in-situ task, an asynchronous execution is preferred when possible, as it allows the GPU-accelerated simulation to make progress on the GPUs while a subset of CPU cores can concurrently execute the in-situ task. 
Our results of NEKO with image generation confirm this hypothesis, however, when using more balanced applications that try to keep both the CPU and GPU equally busy, the situation might change. This is however quite a rare situation these days. 

The QE example with asynchronous in-situ tasks also shows the potential to use idle CPU cores on GPU nodes to decrease the overhead from serial tasks, which could only be executed by a single MPI rank. 
Compared to QE with lossless compression, NEKO with lossy and lossless compression has the lossy compression reusing functions from NEKO and thus deeply coupled with the simulation. It shows that using the idle CPU cores on GPU nodes to execute only the part of the independent in-situ task is beneficial. 
Data compression integrated into both NEKO and QE proves that, as a special in-situ task, asynchronous data compression would only bring small overhead and can reduce the amount of stored data. This allows the researchers to store more frequently simulation results and to have more chance in research discovery. This also can facilitate checkpointing, which is critical for long runs on HPC systems. 
Our previous work~\cite{ju2022understanding} concluded that, from the performance perspective, the asynchronous approach could benefit in-situ tasks with poor scalability while computationally cheap in-situ tasks would prefer a synchronous approach. Thanks to the high peak performance of GPUs, GPU-accelerated simulations typically take much shorter time than CPU-based ones. This also makes computationally cheap in-situ tasks, such as data compression on CPUs, comparably larger and thus these could benefit from asynchronous execution. 
Alternatively, they could be ported to GPUs and execute there, which however might create additional porting difficulties. 

In future work, we will apply the in-situ techniques on Artificial Intelligence (AI) applications which also take advantage of the high computational capacity of the HPC system. Integrating the pre-processing as one in-situ task to the AI training facilities the optimization of pre-processing. 
With the in-situ techniques together with dynamic computational resources, it is possible to further optimize resource usage efficiency. Adding in-situ tasks to rigid applications can also provide additional use cases for malleability with less modification in the original application. 
We will study in-situ techniques integrated into more real-world applications on different HPC systems to build a performance model of in-situ techniques. This can provide general information and help to choose which in-situ technique to use in specific circumstances. 

\section*{Acknowledgment}
This work is partially funded by the “Adaptive multi-tier intelligent data manager for Exascale (ADMIRE)” project, which is funded by the European Union's Horizon 2020 JTI-EuroHPC research and innovation program under grant Agreement number: 956748.
The authors would like to express their gratitude to the Max Planck Computing and Data Facility for providing computing time on the Raven Supercomputer.

\bibliographystyle{IEEEtranS}
\bibliography{refs}

\vspace{12pt}
\end{document}